\documentclass[%
 reprint,
superscriptaddress,
 amsmath,amssymb,
 aps,
]{revtex4-2}

\usepackage[section]{placeins}
\usepackage[columnwise]{lineno}

\usepackage[colorlinks,bookmarks=false,citecolor=magenta,linkcolor=magenta,urlcolor=magenta]{hyperref}
\usepackage{float}
\usepackage{graphicx}
\usepackage{dcolumn}
\usepackage{bm}
\usepackage{epstopdf}
\usepackage{xcolor}
\usepackage{soul}

\usepackage[final]{changes}
\definechangesauthor[name={Peng}, color=cyan]{PENG}
\definechangesauthor[name={Xinhui Ruan}, color=blue]{rxh}

\begin{document}	
	\preprint{APS/123-QED}
	\title{Dynamics and Resonance Fluorescence from a Superconducting Artificial Atom Doubly Driven by Quantized and Classical Fields}
	\author{Xinhui Ruan}
	\affiliation{Department of Automation, Tsinghua University, Beijing 100084, P. R. China}
	\affiliation{Key Laboratory of Low-Dimensional Quantum Structures and Quantum Control of Ministry of Education, Department of Physics and Synergetic Innovation Center of Quantum Effects and Applications, Hunan Normal University, Changsha 410081, China}%
    \author{Jia-Heng Wang}
    \affiliation{School of Integrated Circuits,Tsinghua University, Beijing 100084, China}
    \author{Dong He}
    \affiliation{Key Laboratory of Low-Dimensional Quantum Structures and Quantum Control of Ministry of Education, Department of Physics and Synergetic Innovation Center of Quantum Effects and Applications, Hunan Normal University, Changsha 410081, China}
    \author{Pengtao Song}
    \affiliation{Institute of Physics, Chinese Academy of Sciences, Beijing 100190, China}
    \affiliation{School of Physical Sciences, University of Chinese Academy of Sciences, Beijing 100190, China}
    \author{Shengyong Li}
    \affiliation{Department of Automation, Tsinghua University, Beijing 100084, P. R. China}
    \author{Qianchuan Zhao}
    \affiliation{Department of Automation, Tsinghua University, Beijing 100084, P. R. China}
    \author{L.M. Kuang}
    \affiliation{Key Laboratory of Low-Dimensional Quantum Structures and Quantum Control of Ministry of Education, Department of Physics and Synergetic Innovation Center of Quantum Effects and Applications, Hunan Normal University, Changsha 410081, China}%
    \author{Jaw-Shen Tsai}
    \affiliation{Center for Quantum Computing, RIKEN, Saitama 351–0198, Japan}
    \affiliation{Graduate School of Science, Tokyo University of Science, 1–3 Kagurazaka, Shinjuku, Tokyo 162–0825, Japan}
    \author{Chang-Ling Zou}
    \affiliation{CAS Key Laboratory of Quantum Information, University of Science and Technology of China, Hefei, Anhui 230026, China}   
        \affiliation{Hefei National Laboratory, Hefei 230088, China}
    \author{Jing Zhang}
    \affiliation{School of Automation Science and Engineering, Xi’an Jiaotong University, Xi’an 710049, China}
    \affiliation{MOE Key Lab for Intelligent Networks and Network Security, Xi’an Jiaotong University, Xi’an 710049, China}
    \author{Dongning Zheng}
    \affiliation{Institute of Physics, Chinese Academy of Sciences, Beijing 100190, China}
    \affiliation{School of Physical Sciences, University of Chinese Academy of Sciences, Beijing 100190, China}
    \affiliation{Hefei National Laboratory, Hefei 230088, China}
    \author{O.V. Astafiev}
    \affiliation{Skolkovo Institute of Science and Technology, Nobel str. 3, Moscow, 143026, Russia}
    \affiliation{Moscow Institute of Physics and Technology, Institutskiy Pereulok 9, Dolgoprudny 141701, Russia}
    \affiliation{Royal Holloway, University of London, Egham Surrey TW20 0EX, United Kingdom}
	\author{Yu-xi Liu}
    \affiliation{School of Integrated Circuits,Tsinghua University, Beijing 100084, China}
	\author{Zhihui Peng}%
	\email{zhihui.peng@hunnu.edu.cn}
	\affiliation{Key Laboratory of Low-Dimensional Quantum Structures and Quantum Control of Ministry of Education, Department of Physics and Synergetic Innovation Center of Quantum Effects and Applications, Hunan Normal University, Changsha 410081, China}%
    \affiliation{Hefei National Laboratory, Hefei 230088, China}
	
	\date{\today}

	\begin{abstract}
	We report an experimental demonstration of resonance fluorescence in a two-level superconducting artificial atom under two driving fields coupled to a detuned cavity. One of the fields is classical and the other is varied from quantum (vacuum fluctuations) to classical one by controlling the photon number inside the cavity. The device consists of a transmon qubit strongly coupled to a one-dimensional transmission line and a coplanar waveguide resonator. We observe a sideband anti-crossing and asymmetry in the emission spectra of the system through a one-dimensional transmission line, which is fundamentally different from the weak coupling case. By changing the photon number inside the cavity, the emission spectrum of our doubly driven system approaches to the case when the atom is driven by two classical bichromatic fields. We also measure the dynamical evolution of the system through the transmission line and study the properties of the first-order correlation function, Rabi oscillations and energy relaxation in the system. The study of resonance fluorescence from an atom driven by two fields promotes understanding decoherence in superconducting quantum circuits and may find applications in superconducting quantum computing and quantum networks.
	
	\end{abstract}
	\maketitle
	
	
	\section{Introduction}
Resonance fluorescence is one of the most fundamental physical phenomenon, and comes from the interaction between two-level atoms and classical fields. Early discussions of resonance fluorescence were based on a natural atom and a single color classical driving field. Under strong resonant drive, the emission spectrum from a two-level atom shows Mollow triplet~\cite{Mollow_1969}, which was first observed in natural atoms~\cite{Schuda_1974,Wu1975} and then found in other systems such as trapped ions~\cite{Neuhauser1978,Leibfried2003} and molecules~\cite{Wrigge2008}. It has been experimentally shown that resonance fluorescence is associated with non-classical properties of light, i.e. photon antibunching~\cite{Carmichael_1976,Kimble_1977}. With the advancement of quantum technology, various quantum artificial systems such as quantum dots~\cite{Muller_2007,Xu_2007,NickVamivakas_2009} and superconducting artificial atoms~\cite{Gu_2017,Baur_2009,Astafiev_2010a,Hoi_2012} have been extensively utilized to study resonance fluorescence and the non-classical property of photon antibunching~\cite{Hoi_2012,Lang_2011,Hanschke_2020,Masters2023}. Applications of resonance fluorescence include the detection of quantum states~\cite{Leibfried2003,Albertinale2021}, sensing for absolute calibration of power ~\cite{Honigl-Decrinis_2020} and multiplexed photon counting~\cite{Essig2021}. Recent experimental studies have also unveiled the quantum entanglement features of resonance fluorescence, and showing its potential in entangled photon sources~\cite{Masters2023,Carreno2023}.

The interaction between a cavity and an atom presents another most fundamental model in quantum optics and serves as a basic building blocks in various quantum applications. The introduction of a cavity, or more generally the modification of the environment photonic density of states, offers valuable means to control the properties of atoms and contributes to the development of quantum technologies. When a two-level atom is strongly coupled to a single-mode cavity, the emitted photon by the atom can be stored in the cavity and transferred back to the atom. In the time domain, such strong coupling leads to the coherent  energy exchange between the atom and the cavity, and exhibits Rabi oscillations. In the frequency domain, the atom energy levels are dressed by the lowest Fock states of the cavity, resulting in vacuum Rabi splitting~\cite{Scully_1997}. When the coupling strength is not strong enough, the spontaneous emission rate of the atom could be modified by the cavity, which is known as the Purcell effect~\cite{Purcell_1946}. 

Dressing of atoms by both classical driving fields and quantized cavity vacuum fields significantly alters atom properties, motivating investigations into doubly-dressed atoms by both driving and cavity fields from both fundamental and applied perspectives. The modulation of resonance fluorescence in a system, where an atom is resonantly coupled to a single-mode cavity has been theoretically~\cite{Savage_1989} and experimentally~\cite{Peng_2022} demonstrated. The linewidth of sidebands in this system is determined by the coherent coupling strength between the atom and the cavity, rather than the atom's decay rate. The presence of a cavity mode when an atom is detuned with a single-mode cavity results in the modification of a single fluorescence sideband~\cite{Quang_1993,Roy_2012}. In the strong coupling regime, each Mollow triplet splits into multiplets and the emission spectrum becomes asymmetric~\cite{Freedhoff_1993}. In the weak coupling regime, only the sideband emission enhancement effect was experimentally observed in the quantum dot system~\cite{Kim_2014}. Predictions also suggest that the fluorescence linewidth can be reduced below the vacuum level in a squeezed vacuum field environment~\cite{Toyli_2016,Carmichael_1987}, and complete fluorescence elimination can be achieved with bichromatic laser fields~\cite{He_2015, Zhu_1996a, Ficek_1999, He_2019}. 

Superconducting qubits, among various natural and artificial atoms, offer a unique experimental platform for studying fundamental light-atom interactions. With the progress in superconducting quantum circuits~\cite{Gu_2017, Kwon_2021, Blais_2021}, investigating quantum optics on-chip using superconducting artificial atoms has become feasible by appropriate design circuit parameters. Experimental explorations of resonance fluorescence from strongly coupled artificial atoms to one-dimensional open space~\cite{Astafiev_2010a, Hoi_2012} and chiral artificial atoms~\cite{Joshi2023} have been conducted. Here, we study a system that the atom formed by the transmon qubit is strongly coupled to a single-mode cavity and a transmission line simultaneously. First, we observe the resonance fluorescence of the atom-cavity coupled system, which has multiplet peaks and explain it with doubly dressed-state theory. The spectrum here is different from the Mollow triplet and has splitting in sidebands when the driving strength is equal to the detuning between the atom and the cavity. This is caused by the strong coupling between them. Moreover, the relative amplititudes of the sidebands are asymmetric.

We use time-domain measurement to study the dynamic properties of the system, including first-order correlation function, Rabi oscillations and spontaneous emission process. We find that these processes are modulated by a classical driving field and a cavity field simultaneously. In order to spatially separate the atom's drive and output signals, the drive of the atom is added from the cavity port. Through applied measurement, we find that our driving method only affects the coherent properties of the atom, but does not affect its incoherent properties. 
Finally, we study the influence of photons inside the cavity on the resonance fluorescence spectrum by pumping the cavity and compare the results with the case, when the cavity is in the vacuum. When the photon number inside the cavity becomes so large such that the splitting sub-levels overlap, we find that the resonance fluorescence spectra approach to the results in Ref.~\cite{He_2015} in which the atom is driven by two classical driving fields. 

\section{Experimental setup and model}
	\begin{figure*}[t]
		\includegraphics[scale=0.8]{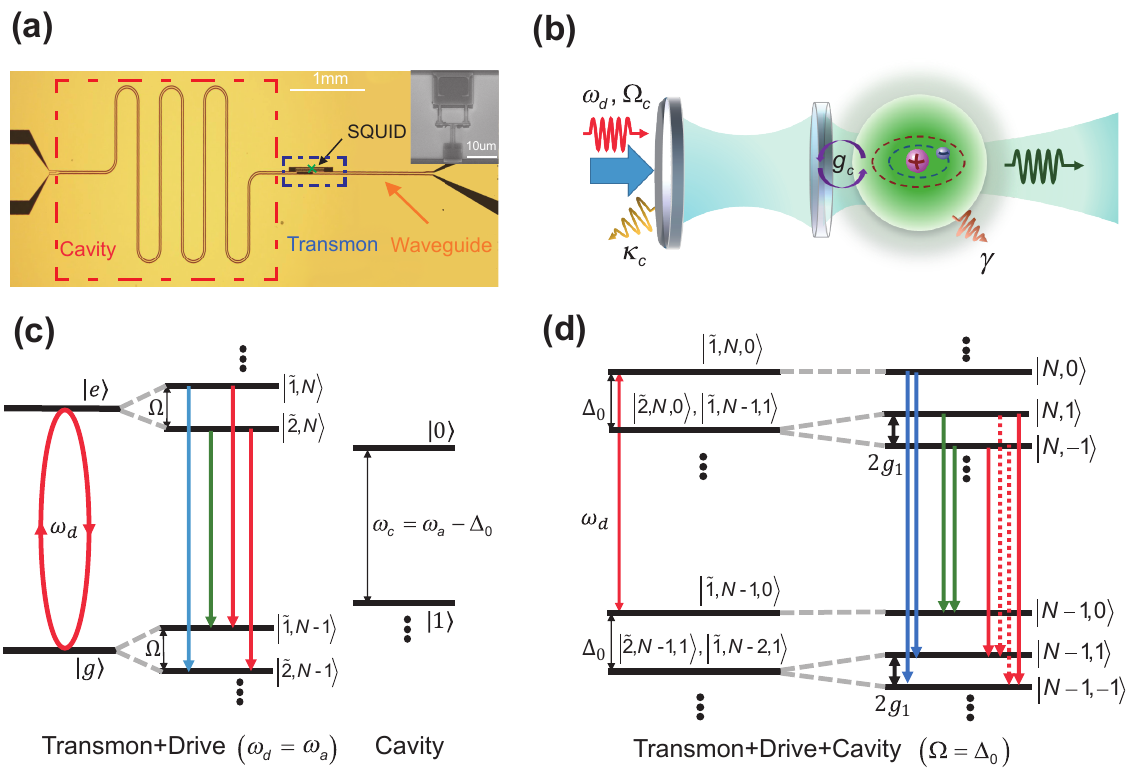}
		\caption{(a) Optical image of the fabricated device. The cavity is a coplanar waveguide (CPW) resonator. The transmon qubit acts as an artificial atom with tunable transition frequency. The insert shows the scanning electron microscopy of the SQUID. (b) A system with an atom coupled with a cavity and a waveguide simultaneously. The cavity acts as an engineered environment of the atom and can affect the fluorescence spectrum of the atom. (c) The energy level diagram for drive dressed states of the transmon and the ground states of the cavity mode. The cavity mode frequency $\Delta_0$ is red detuned to the transition frequency of the transmon, and coupling between the cavity and the dressed transmon is suppressed due to detuning. (d) The energy level diagram for the doubly dressed system, when $\Omega=\Delta_0$. The colored lines with arrows show the allowed transitions.}
      \label{figure1}
	\end{figure*}
The device is composed of a half-wavelength one-dimensional (1D) coplanar waveguide (CPW) resonator and a transmon qubit, which are strongly coupled through a capacitor as shown in Fig.~{\ref{figure1}(a)}. Besides, the transmon is strongly coupled to a 1D waveguide simultaneously. The transmon has a SQUID loop which is used to tune its transition frequency from $2\pi\times7.5\,$GHz to lower ones. Because the  anharmonicity is around $2\pi\times625\,$MHz, the transmon can be treated as a two-level system with ground state $\left|g\right\rangle$ and first excited state $\left|e\right\rangle$. In this experiment, we fix the transition frequency of the transmon to $\omega_a/2\pi=6.814\,$GHz by adjusting the flux in the SQUID loop. The dephasing rate of the transmon at $ \omega_a $ is $ \gamma_2=\gamma_1/2+\gamma_\varphi=2\pi\times2.8\,$MHz, where $\gamma_1$ is the energy relaxation rate and $\gamma_\varphi$ is the pure dephasing rate. $\gamma_1=\gamma_e+\gamma_n$ is formed by the ralaxation rate through the transmission line $\gamma_e$ and the nonradiative relaxation rate to the environment $\gamma_n$. As the the emission efficiency of the transmon through the transmission line is $ \eta \equiv\gamma_e/2\gamma_2=62\% $, $ \gamma_e/2\pi=3.5\,$MHz, which helps with the collection of the emitted signals from the transmon.
The frequency of CPW resonator is $ \omega_c/2\pi=6.777\,$GHz and thus the red detuning with the transmon is $ \Delta_0/2\pi=37.0\,$MHz. The decay rate of the resonator is $ \kappa/2\pi=1.5\,$MHz. The coupling strength between the resonator and the transmon is $ g_c/2\pi=7.5\pm1\,$MHz. All related detailed information of these parameters is given in Appendix~\ref{parameters}.

Our system can be schematically described as shown in Fig.~{\ref{figure1}(b)}. The driving field for the two-level atom is applied through the cavity at frequency $ \omega_d $ with driving strength $ \Omega_c $. The full Hamiltonian of the system with the rotating wave approximation of $\omega_d$ is
    \begin{align}\label{eqn-0}
	H=\Delta_c a^{+} a+\frac{\Delta_a}{2} \sigma_z+g_c\left(a^{\dag} \sigma_{-}+a \sigma_{+}\right)+\frac{\Omega_c}{2}\left(a+a^{+}\right),
    \end{align}
with $\Delta_{a/c}=\omega_{a/c}-\omega_d$. $a$ is the annihilation operator of cavity mode, $\sigma_{-}=\left|g\right\rangle\left\langle e \right|$ and $\sigma_{+}=\left|e\right\rangle\left\langle g \right|$ are the ladder operators of the atom and $\sigma_z=|e\rangle \langle e|-|g\rangle \langle g|$ is the population operator of atom. In addition to coupling with external driving field and cavity field, the atom also couples with the continuum modes of a waveguide. The corresponding system dynamics could be described by the Master equation 
\begin{align}\label{6}
\dot{\rho }=-i[H,\rho]+\frac{\gamma_1}{2}D\left( {{\sigma }_{-}} \right)\rho+\gamma_\varphi D\left( {{\sigma }_{z}} \right)\rho+\frac{\kappa}{2} D(a)\rho, 
\end{align}
with $D(A)\rho =2A\rho {{A}^{\dagger }}-{{A}^{\dagger }}A\rho -\rho {{A}^{\dagger }}A$ is the Lindbald operator, with the qubit energy relaxation rate $\gamma_1=3.6\,$MHz and pure dephasing rate $\gamma_\varphi=1.0\,$MHz.  Output signals from the atom are collected directly through the waveguide, which could suppress the influence of the strong input drive on the output signals when the driving field is applied through the cavity side, because the cavity acts as a filter in the observation of the atom spectra due to $  |\omega_d-\omega_c |\gg\kappa $. Therefore, according to the input-output theory, the collected signal could be obtained as $b_{out}=\sqrt{\gamma_e/2}\sigma_{-}$. 

Under an external strong drivinging field, the cavity field could be approximated as a sum of a classical field and a weak cavity mode excitation as $a\rightarrow \alpha+a$, with the steady-state drive field  $\alpha=-\Omega_c/2\Delta_c$, then the Hamiltonian becomes 
\begin{align}      
        H=\Delta_c a^{+} a+\frac{\Delta_a}{2} \sigma_z+g_c\left(a^{\dag} \sigma_{-}+a \sigma_{+}\right)+g_c\alpha\left(\sigma_{-}+\sigma_{+}\right).
        \label{eqn-1}
\end{align}
Consequently, the model can be divided into two parts, the conventional atom-cavity photon coupling and the strong drive with the Rabi frequency $\Omega=2g_c\alpha$ applied to the atom.  In our experiments, the drive is resonant with the atom  $\omega_d=\omega_a$ and the cavity mode is largely detuned, thus $\Delta_a=0$ and  $\Delta_c=-\Delta_0=\omega_c-\omega_d$.  The transition from Eq.~(\ref{eqn-0}) to  Eq.~(\ref{eqn-1}) can be understood in the way that the cavity field is divided into the classical and fluctuation parts. These two equations are equivalent for the calculation of the steady-state fluorescence spectrum, because they are consistent in the description of photon fluctuations. But when studying the time evolution of the system, the influence of classical coherent photons can not be omitted as we show in Section~\ref{evolution}.  Note, the mean photon number $\langle n\rangle$ we mention below refers to the fluctuation photons  in the cavity.

From the Hamiltonian in Eq.~(\ref{eqn-1}), our system provides a versatile platform to investigate the dynamics of dressed atoms. When the cavity mode is largely off-resonant with the atom, the model is equivalent to the conventional fluorescence resonance. Fig.~\ref{figure1}(c) illustrates the energy diagram, with the driving field dressed atom states, which are expressed as 
\begin{align} 
|\tilde i,N\rangle  = \frac{1}{{\sqrt 2 }}\left[ {|g,N\rangle- {{\left( { - 1} \right)}^{ i}}|e,N-1\rangle} \right],\,i=\{1,2\},
\label{eqn-2}
\end{align}
with eigenenergy $E_{\tilde i,N}=N\omega_d- {{\left( { - 1} \right)}^{ i}}\Omega/2$ for $N$ photons in the driving field. The cavity Fock state transitions are red-detuned from dressed-state transitions, and their couplings are neglected. There are three different transition frequencies $\omega_d$ and $\omega_d\pm\Omega$ between the dressed states, manifesting the Mollow triplet emission spectrum in fluorescence resonance. 

When the dressed-state transition near-resonant with the cavity mode, i.e., driving strength $\Omega=\Delta_0$, the energies of $|\tilde 2,N,n\rangle$ and $|\tilde 1,N-1,n+1\rangle$ are degenerate. Here, $|\tilde i,N,n\rangle \equiv |\tilde i,N\rangle \otimes |n\rangle$ with $|\tilde i,N\rangle$ given in Eq.~{(\ref{eqn-2})} and $|n\rangle$ the Fock state of the cavity field. In this case, the driving field dressed atom is further dressed by the cavity Fock states, resulting in doubly dressed states
	\begin{align}\label{eqn-3}
\left| N,\pm n \right\rangle =\frac{1}{\sqrt{2}}\left[ |\tilde{1},N-n,n\rangle \mp |\tilde{2},N-n+1,n-1\rangle  \right],
    \end{align}
 and the corresponding eigenenergy is $E_{N,\pm n}=N\omega_d-\left(2n-1\right)\Omega/2\pm g_1\sqrt n$, with $g_1=g_c/2$. In Fig.~{\ref{figure1}(d)}, we show the energy levels of the whole system with $\Omega=\Delta_0$, when the mean photon numbers $\langle n\rangle$ of the cavity is less than one. Seven peaks are expected in the emission spectrum for this situation, which can only be resolved in the case of $g_1>\kappa,\gamma$. When increasing the cavity excitation with $\langle n\rangle\gg1$, this model approaches the two classical driving fields applied to the atom, and there will be more emission peaks as observed in Ref.~\cite{He_2015}.

In this work, we experimentally investigate the modified properties of classical driving field dressed atom by introducing the cavity mode, either in vacuum or classical field. We also note that our system could be turned into a significant different model, by setting the atom-cavity detuning to zero. In this case, the atom is directly dressed by the vacuum cavity field as the system is in the strong coupling regime, and the vacuum-dressed atom could be further dressed by a classical driving field. Such a different model was investigated in Ref.~\cite{Peng_2022}, the anomalous side-peaks broadening in the fluorescence spectrum arises from relaxation transitions of the dressed states when the atom perturbs the driven cavity.

	\section{Resonance fluorescence of doubly dressed states}

    \begin{figure}[h] 
		\includegraphics[scale=0.68]{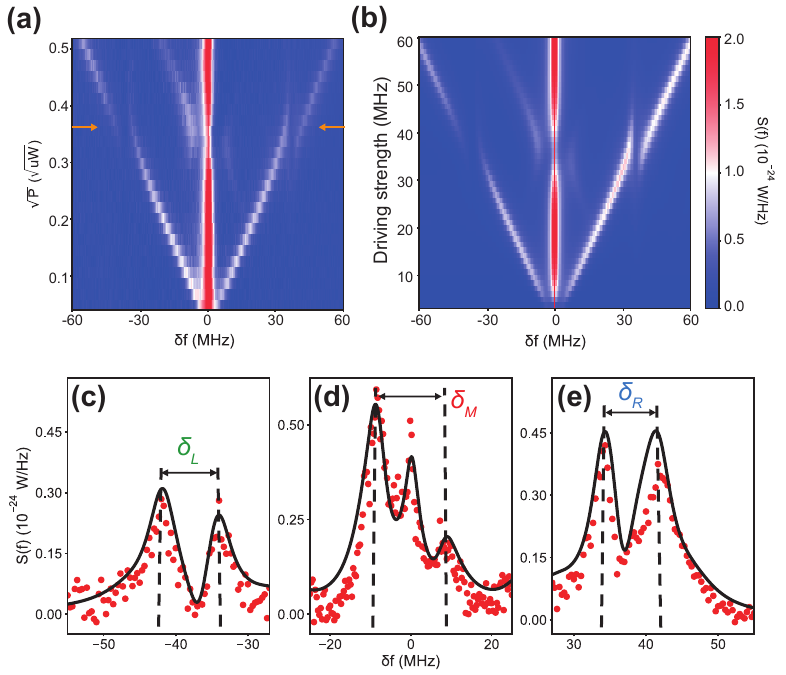}
        \caption{(a) Resonance fluorescence spectra of the transmon coupled to a CPW cavity against $\sqrt{P}$, where $P$ is driving power at a room temperature. The driving frequency $\omega_d$ is in resonance with the transmon. $\delta f=f - \omega_d/2\pi$ is the frequency detuning from the driving frequency. (b) A theoretical plot of the emission spectra obtained by solving the master equation. Measured emission spectrum around (c) $-\Delta_0/2\pi$, (d) 0 and (e) $\Delta_0/2\pi$ for the situation that $\Omega=\Delta_0$, which is also marked by gold arrows in (a). The red dots in (c)-(e) are the experimental data, while the solid lines are the theoretical simulation results.}
        \label{figure2}
    \end{figure}
 
Figure~{\ref{figure2}(a)} shows change of the resonance fluorescence spectra with the driving amplitudes $\sqrt{P}$ generated from an arbitrary waveform generator (AWG). We can measure the spectra using data acquisition card with the measurement setup shown in Appendix~{\ref{Asec3:level3}}. The total attenuation on the input line is about 103 dB. The spectra show a linear dependence of $\sqrt{P}$ for the distance between the central peak and the side peaks, agreeing with the behavior of conventional resonance fluorescence. 

When one of the side peak frequency is near resonance with the cavity, two anti-crossings in the sidebands appear, as shown in Fig.~\ref{figure2}(a) with $\sqrt{P}$ around the point marked by the gold arrow, which corresponds to $\Omega=\Delta_0$. The anti-crossings are the indisputable evidence that there is strongly resonant coupling between a sideband of Mollow Triplet and a cavity, which agrees with our prediction shown in Fig.~\ref{figure1}(d) and indicates the double-dressing of the atoms by the classical driving field and the vacuum cavity field. We have to point out that our results are fundamentally different from the weak resonant coupling between a sideband of Mollow Triplet and a cavity in Ref.~\cite{Kim_2014}, where the coupling strength $g/2\pi=15.3\,$GHz is smaller than the decay rate of the cavity $\kappa/2\pi=36\,$GHz. In the weak coupling regime, only the atom emission rate could be significantly affected, instead of the splitted emission spectrum due to the cavity dressed states. Therefore, only the enhancement of resonant sideband relative to the detuned sideband in the resonance fluorescence spectra is observed in Ref.~\cite{Kim_2014}, and the avoided crossing of resonance fluorescence has yet to be demonstrated.

To better understand the spectra, we perform theoretical simulations by using the Master equation in Eq.~(\ref{6}). The output field of the transmon qubit in the 1D waveguide is \cite{Astafiev_2010a}
\begin{align}\label{eqn-4}
	I(x,t)=i\frac{\hbar {{\gamma_1 }}}{{{\phi }_{p}}}\left\langle {{\sigma }_{-}(t)} \right\rangle {{e}^{ik\left| x \right|-i\omega t}},
\end{align}
with the qubit placed at the position of $x$=0. ${{\phi }_{p}}=\hbar \Omega /{{I}_{0}}$ is the dipole moment and $I_0$ is the amplitude of the input field. The definition of stationary emission spectrum of the atom is 
\begin{align}\label{eqn-5}
    S(\omega )={{\pi }^{-1}}{{\lim }_{t\to \infty }}\operatorname{Re}\int_{0}^{\infty }{\left\langle {{\sigma }_{+}}(t){{\sigma }_{-}}(t+\tau ) \right\rangle }{{e}^{-i\omega \tau }}d\tau,
  \end{align}
where $t\to \infty $ means that the state used here is the steady-state. The results are shown in Fig.~{\ref{figure2}(b)}. We can see that the simulations agree well with the experimental results. The detailed comparison between simulations and the experiment at doubly dressed states is shown in Figs.~{\ref{figure2}(c)-(e)}. We remove coherent Rayleigh-scattered radiation~\cite{Lang_2011} and correlated noise between the two output lines of the hybrid coupler~\cite{Marcus_2010}, while preserve incoherent emission spectrum. 
The frequency splittings of left (Fig.~{\ref{figure2}(c)}) and right (Fig.~{\ref{figure2}(e)}) sidebands are $ \delta_L= \delta_R=8\,\text{MHz}\approx2g_1/2\pi $, corresponding to the green and blue transition lines in Fig.~{\ref{figure1}(c)}. As shown in Fig.~{\ref{figure2}(d)}, around the center frequency, there are three peaks and the distance between side peaks is $\delta_M=16\,\text{MHz}\approx4g_1/2\pi$, corresponding to the red transition lines in Fig.~{\ref{figure1}(c)}. Note that the peak heights in  Figs.~{\ref{figure2}(c)} and \ref{figure2}(e) are asymmetric, which can be explained by different population distributions of related dressed states. Combined with Fig.~{\ref{figure1}(c)}, we can see that the right sidebands are caused by the emissions from $\left| N,0 \right\rangle$ while the left sidebands are from $\left| N,\pm 1 \right\rangle$. From  Eq.~(\ref{ddpopulation}) in Appendix~\ref{theory}, we find that the population of   $\left| N,\pm n \right\rangle$ decreases with $n$. Therefore,  the population of state $\left| N,0 \right\rangle$ is greater than that of state $\left| N,\pm 1 \right\rangle$, causing the spectrum intensity around $\delta f=\Delta_0/2\pi$ to be higher than $\delta f=-\Delta_0/2\pi$. 
    
    \begin{figure}[h]
	   \includegraphics[scale=0.47]{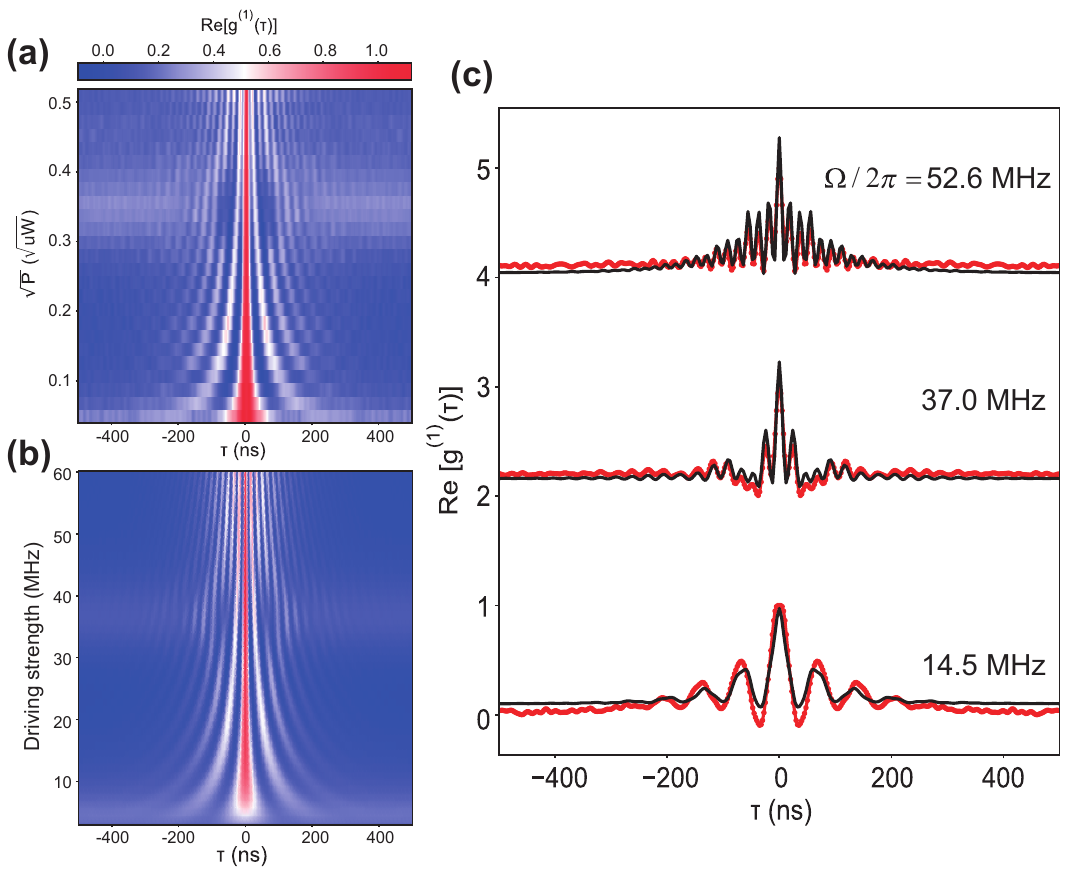}
	   \caption{(a) Measured real part of the two-time correlation function $g^{(1)} (\tau)$ against $\sqrt{P}$. (b) The simulated real part of $g^{(1)} (\tau)$ is shown as a function of driving strength. (c) The contrast of measured (red dots) and calculated (black lines) real parts of $g^{(1)} (\tau)$ with different driving strength $\Omega$. The middle line with $\Omega/2\pi$=37.0 MHz corresponds to the doubly dressed case.}
	\label{figure3}   
    \end{figure}

Eq.~(\ref{eqn-5}) shows that the resonance fluorescence is obtained via the Fourier transfer of the first-order  correlation function $ g^{(1)} (\tau) $ with
    \begin{align}
    {{\text{g}}^{\left( 1 \right)}}\left( \tau  \right)={{\lim }_{t\to \infty }}\left\langle {{\text{ }\!\!\sigma\!\!\text{ }}_{+}}\left( \text{t} \right){{\text{ }\!\!\sigma\!\!\text{ }}_{-}}\text{ }\!\!~\!\!\text{ }\left( \text{t}+\text{ }\!\!\tau\!\!\text{ } \right) \right\rangle,
    \end{align}
which can characterize coherence of this system.
In experiments, we can measure ${{\text{g}}^{\left( 1 \right)}}\left( \tau  \right)$ by correlating the output signals from waveguide, as ${{\text{g}}^{\left( 1 \right)}}\left( \tau  \right)={{\Gamma }^{(1)}}(\tau )-\Gamma _{bg}^{(1)}(\tau )$ with ${{\Gamma }^{(1)}}(\tau )=\sum\nolimits_{t}{I(t)*I(t+\tau )}$ and the correlated noise background $\Gamma _{bg}^{(1)}(\tau )$(experimental details are in Appendix~{\ref{Asec3:level3}}). Here, $I(t)$ is given in Eq.~{(\ref{eqn-4})}. We normalize ${{\text{g}}^{\left( 1 \right)}}\left( \tau  \right)$ by setting ${{\text{g}}^{\left( 1 \right)}}\left( \tau =0 \right)=1$. The traces of ${{ \text{g}}^{\left( 1 \right)}}\left( \tau  \right)$ at different input driving amplitudes $\sqrt{P}$ are shown in Fig.~{\ref{figure3}(a)} and each trace is averaged by $1\times {{10}^{8}}$ times. Fig.~{\ref{figure3}(b)} is the simulation results of ${{\text{g}}^{\left( 1 \right)}}\left( \tau  \right)$. It is shown that the oscillation of ${{\text{g}}^{\left( 1 \right)}}\left( \tau  \right)$ becomes faster when $\sqrt{P}$ increases, which is consistent with results in the frequency domain. We also show the comparison of experiments and simulations at driving strength $\Omega/2 \pi =$14.5, 37.0, 52.6$\,$MHz in Fig.~{\ref{figure3}(c)}, which correspond to the red detuned sideband, resonant and blue detuned sideband with the cavity, respectively. We have to point out that when $\Omega $ is large, the mean thermal photon $\langle{{n}_{th,c}}\rangle$ in cavity is not negligible because strong drive heats up the cavity. Thus, we set $\langle{{n}_{th,c}}\rangle=0.5$ in simulations to $\Omega/2 \pi =$37.0\,\text{MHz}, 52.6$\,$MHz in Fig.~{\ref{figure3}(c)} and find that it fits well with experimental results. As shown in the middle part of Fig.~{\ref{figure3}(c)}, when the driving strength $\Omega=\Delta_0=2\pi\times37.0\,$MHz, the oscillations in measured $g^{(1)} (\tau)$ are modulated by the cavity field and the driving field simultaneously. It is quite different from the oscillations in measured $g^{(1)} (\tau)$ when $\Omega/2\pi=14.5\,$MHz or $52.6\,$MHz in Fig.~{\ref{figure3}(c)}.

    \section{\label{evolution}Dynamic evolution of the Transmon-cavity coupled system}
    \begin{figure*}[t]
	   \includegraphics[scale=0.8]{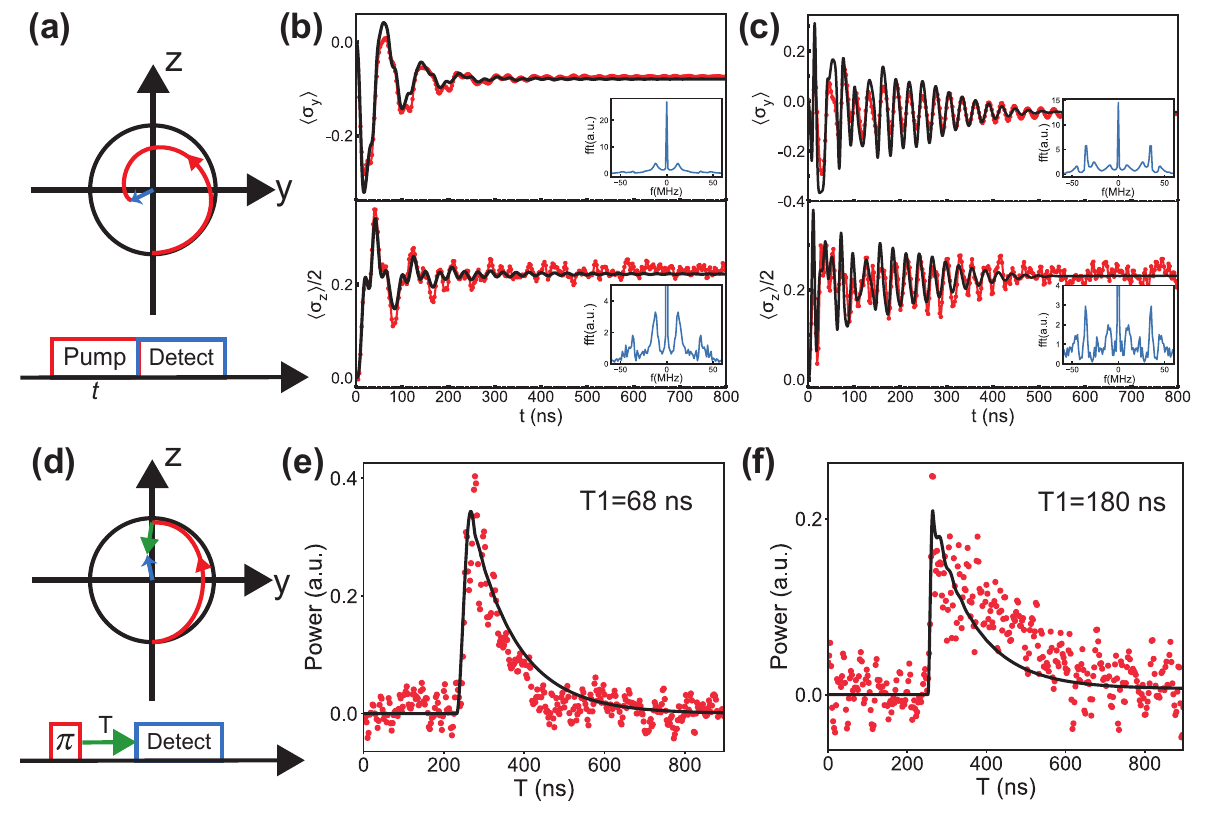}
	   \caption{(a) Pulse sequence of the evolution detection in (b-c) and schematic evolution of the transmon on the y-z cross of Bloch sphere. (b) Time evolution of the quantum states $\langle \sigma_y \rangle$ and $\langle \sigma_z \rangle$ with $\Omega/2\pi$=14.5 MHz. Time interval for each experimental point is 2.5 ns. An insert is Fourier transform of experimental time evolution. (c) Time evolution of $\langle \sigma_y \rangle$ and $\langle \sigma_z \rangle$ with $\Omega/2\pi$=37.0 MHz, corresponding to the doubly dressed situation. (d) Pulse sequence of the power trace detection in (e-f) and the corresponding evolution on the Bloch sphere. (e) Single power trace at $\theta_r=\pi$, corresponding to the state $|e\rangle$ of the Transmon for $\Omega/2\pi$=14.5 MHz. (f) Single power trace at $\theta_r=\pi$ for $\Omega/2\pi$=37.0 MHz. The red dotted lines and red dots here are the experimental data, while the black solid lines are the calculated data using the Master equation.}
    \label{decoherence}
    \end{figure*}

    As we mentioned earlier, Eq.~{(\ref{eqn-0})} and Eq.~{(\ref{eqn-1})} are different for the process before the artificial atom goes into the steady-state, so we use Hamiltonian in Eq.~(\ref{eqn-0}) to simulate the dynamics of the coupled system. The state of qubit can be characterized by measuring the signal output amplitude $I(t)$ and power $P(t)$ through the waveguide, as $I(t)$ is shown in Eq.~(\ref{eqn-4}) and 
    \begin{align}\label{eqn-7}
    P(t)=\left( \frac{\hbar {{\omega }_{a}}{{\gamma_1 }}}{4} \right)\left( 1+\left\langle {{\sigma }_{z}}\left( t \right) \right\rangle  \right){{e}^{-{{\gamma_1 }}t}}.
    \end{align}
    In order to compare the difference of evolution between singly dressed states and doubly  dressed states, we show the evolution of $\left\langle {{\sigma }_{y}} \right\rangle =\operatorname{Im}\left\{ \left\langle {{\sigma }_{-}} \right\rangle  \right\}$ and $\left\langle {{\sigma }_{z}} \right\rangle $ of the qubit from the ground state $\left| g \right\rangle $ to steady-state when $\Omega =2\pi\times14.5\,$MHz and $\Omega =2\pi\times37.0\,$MHz in Figs.~\ref{decoherence}(b) and 4(c), respectively. We change the duration $t$ of the pump pulse to let the qubit evolve to different states on the Bloch sphere as $\left| \psi_q(t) \right\rangle=\cos ({{\theta }_{r}(t)}/2)\left| 0 \right\rangle +\sin ({{\theta }_{r}(t)}/2)\left| 1 \right\rangle $, where ${{\theta }_{r}(t)}$ is the Rabi angle. Meanwhile, we detect the output amplitude and power of the qubit through the waveguide in real time (Fig.~\ref{decoherence}(a)). The population of the qubit oscillates with certain Rabi frequencies and decays before it reaches the steady-state. The decay is caused by the relaxation rate of the system. Unlike the evolution of a qubit resonantly driven by a classical driving field with ${\dot{\theta} }_{r}(t)=\Omega$ or of a qubit coupled to a cavity field~\cite{Mooij_2010,wallraff2004} with ${\dot{\theta} }_{r}(t)=\sqrt{4g_c^2+(\omega_q-\omega_c)^2}$, the Rabi frequencies in our system are complex. From the inserts in Figs.~\ref{decoherence}(b) and \ref{decoherence}(c), we can see that there are more than one oscillation frequency. The high and narrow central peak contains signal of the Rayleigh scattering part. Leaving out the Rabi frequency at $\pm\sqrt{4g_c^2+\Delta_0^2}\approx\pm2\pi\times39.0\,$MHz caused by the coupling with the cavity, the other frequencies are equal to the peak positions of the resonance fluorescence spectrum in Fig.~\ref{figure2}(a) with the same input driving strength. The measurement data has some distortion in Fig. 4(c), caused by the limitation of the sampling rate from our data acquisition card and the low emission efficiency of the qubit. The black solid lines in Figs.~\ref{decoherence}(b) and \ref{decoherence}(c) are fittings using the Master equation with the original Hamitanion in Eq. (\ref{eqn-0}), because the drive also populates the cavity with photons, which affects the evolution of the qubit. 
    
    Next, we compare their different behavior in spontaneous emission process of the qubit. As shown in Fig.~\ref{decoherence}(d), we prepare the qubit into the excited state  $\left| e \right\rangle $ with $\pi$-pluse and measure the free evolution process of energy by collecting the output power of the qubit with different waiting time $T$. 
To obtain energy relaxation time ${{T}_{1}}$ of the qubit, we  fit of the power traces by exponents in Fig.~\ref{decoherence}(e) and \ref{decoherence}(f) with $e^{-T/T_1}$. The energy relaxation time of doubly dressed states ${{T}_{1,D}}=180\,$ns is longer than that of single dressed states ${{T}_{1,S}}=68\,$ns, which is consistent with the change trend of central peak linewidth in Fig. 2(a) (see Appendix \ref{Asec5:level5}). Therefore, an additional modification from the cavity can improve the energy relaxation time of the qubit when $\kappa < \gamma_2$. The solid curves in Figs.~\ref{decoherence}(e) and \ref{decoherence}(f) are simulation results using the Master equation.

    In the past 20 years, there has been significant attention paid to the degradation of coherence or efficiency due to the coupling of a qubit~\cite{Simmonds2004} or a quantum emitter~\cite{Zhou2020} to unwanted degrees of freedom, such as two-level defects or parasitic modes in the environment. Our experiments provide a method to ascertain whether a driven qubit or driven emitter is coupled to unwanted degrees of freedom in the environment. As shown in Figs.~\ref{decoherence}(b) and (c), we reveal the dynamics of a driven qubit coupled to a cavity by time-domain measurements. The cavity mode can be considered as an unwanted degrees of freedom in the environment. In Fig.~\ref{decoherence}(b), the driven qubit is still detuned from the cavity and the qubit shows more like free evolution process. There is a typical Mollow Triplet structure shown in the inserts. In Fig.~\ref{decoherence}(c), the driven qubit is strongly resonant to the cavity and the dynamics has to be described by the driven qubit-cavity coupled system. The qubit is strongly modulated by the cavity and the spectrum structure changed drastically as shown in the inserts. In future experiments, it would be advantageous to avoid coupling driven qubits or emitters to unwanted degrees of freedom such as two-level defects in the environment. 

\section{Doubly-Dressed Resonance fluorescence for non-vacuum state in a cavity}
    In the above experiments, the mean photon number in the cavity is close to zero. However, with some photons in the cavity, the resonance fluorescence spectra of the qubit-cavity coupled system will change because more energy levels of the doubly dressed states need to be taken into account. 

\begin{figure*}[t]
	\includegraphics[scale=1.7]{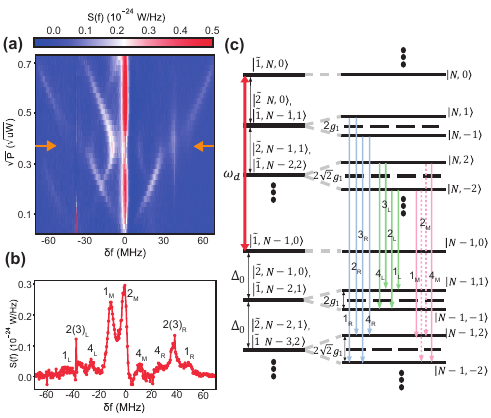}
	\caption{(a) The resonance fluorescence spectra of a qubit against driving amplitude $\sqrt{P}$, when there is one photon stored in the cavity. (b) The corresponding spectrum of doubly dressed states marked by orange arrows in (a). The $\delta f$ of the marked peaks are $\delta {{f}_{1L,\ 1R}}=\pm 47\ \text{MHz}$, $\delta {{f}_{4L,\ 4R}}=\pm 27\ \text{MHz}$ and $\delta {{f}_{1M,\ 4M}}=\pm 11\ \text{MHz}$.  (c) The energy levels and related transitions of the doubly dressed states with one photon in cavity.}
     \label{figure5}
\end{figure*}

    In Fig.~\ref{figure5}(a), we apply an additional continuous pump with the frequency $\omega_{pc}$ that resonantly populates the cavity. Besides, the driving field with the amplititude $\sqrt{P}$ and the frequency $\omega_d$ is resonantly applied to the qubit through the cavity. The amplitude of $\omega_{pc}$ is designed to keep the mean photon number $\langle n\rangle$ in the cavity to be 1.4. With $\langle n\rangle>1$, the states $\left| N,\pm 2 \right\rangle $ will be present. As shown in Fig.~\ref{figure5}(c), the energy spacing between $\left| N,2 \right\rangle $ and $\left| N,-2 \right\rangle $ is $2\sqrt{2}{{\text{g}}_{1}}$, which is not equal to the space between $\left| N, 1 \right\rangle $ and $\left| N, -1 \right\rangle$. The new transitions $\left| N,\pm 1 \right\rangle\rightarrow \left| N-1,\pm 2 \right\rangle$,  $\left| N,\pm 2 \right\rangle\rightarrow \left| N-1,\pm 1 \right\rangle $ and $\left| N,\pm 2 \right\rangle\rightarrow \left| N-1,\pm 2 \right\rangle $ will generate new peaks in the emission spectrum, which are marked by lines with arrows and numbers with subscript in Fig. 5(c). These new peaks are located at frequencies ${{\omega }_{d}}\pm {{\Delta }_{0}}\pm (\sqrt{2}\pm 1){{g}_{1}}$, ${{\omega }_{d}}$ and ${{\omega }_{d}}\pm 2\sqrt{2}{{g}_{1}}$. However, for peaks ${{2}_{L,R}}$ and ${{3}_{L,R}}$, the frequency spacing from ${{\omega }_{d}}\pm {{\Delta }_{0}}$ is $(\sqrt{2}-1){{g}_{1}}\approx 2\pi\times1.6\, $MHz, which is smaller than the linewidth of qubit, so they overlap with each other and form the peaks at ${{\omega }_{d}}\pm {{\Delta }_{0}}$. Therefore, we observe the resonance fluorescence from an atom driven by two classical fields which is similar to the results reported in Ref.~\cite{He_2015}. We show the corresponding measured spectrum of the above analysis in Fig. 5(b). When the driving amplitude of the qubit is equal to $\Delta_0$, we can see $2_\text{M}$,  ${{1}_{\text{L,R,M}}}$ and ${{4}_{\text{L,R,M}}}$ in the spectrum. The spacings of these peaks from ${{\omega }_{d}}\pm {{\Delta }_{0}}$ and ${{\omega }_{d}}$ become larger compared to the results in Fig. 2 and reach $2\pi\times$10 MHz$\left( \approx \left( \sqrt{2}+1 \right){{g}_{1}} \right)$ and $2\pi\times$11 MHz$\left( \approx 2\sqrt{2}{{g}_{1}} \right)$. We attribute the peak at $\delta f= -\Delta_0$ to the four-wave mixing with the artificial atom~\cite{Dmitriev2017}. Overall, the spectrum here is very close to that of two classical driven fields in Ref.~\cite{He_2015} which has nine peaks. When $\langle n\rangle\gg1$, the frequencies of $(1,2,3,4)_\text{L,R}$ would be ${\omega }_{d}\pm {{\Delta }_{0}}\pm (\sqrt{n+1}\pm\sqrt{n}){g}_{1}$ with $(\sqrt{n+1}+\sqrt{n}){g}_{1}\approx 2\sqrt{n} {g}_{1}$ and $(\sqrt{n+1}-\sqrt{n}){g}_{1}\approx 0$. As a result, the cavity field can be equivalent to a classical field with driving strength $\Omega_c=2\sqrt{n} {g}_{1}$. For our system, this assumption holds when $\langle n\rangle\geq1$. 

    From Fig.~\ref{figure2} and Fig.~\ref{figure5}, we can observe the resonance fluorescence crossover of an atom from being driven by a quantum vacuum field and a classical field to being driven by two classical fields through populating the photons inside the cavity. The most essential difference in the two cases is that the non-uniform energy level spacing of the doubly dressed states caused by the coupling between the cavity and the atom and depends on the mean photon number inside the cavity. The coupling strength $(\sqrt{n+1}-\sqrt{n}){g}_{1}$ will bring more peaks and the existence of states with the cavity in vacuum field will cause anti-crossing in the emission spectrum when it is larger than other dissipation rates in the system. However, when $(\sqrt{n+1}-\sqrt{n}){g}_{1}$ is smaller than other dissipation rates in the system, it shows resonance fluorescence from an atom driven by two classical fields. 

\section{Conclusion}
    In conclusion, we demonstrate the resonance fluorescence of a transmon qubit driven by one classical field and one quantum vacuum field of a cavity. The cavity acts as a modulator of the resonance fluorescence and can induce sidebands with asymmetric intensity of the spectrum, which has been verified in Ref.~\cite{Kim_2014}. We observe fundamentally different resonance fluorescence that there are anti-crossings in the emission spectrum because the sideband of Mollow Triplet interacts with the cavity in a strong coupling regime. The results in our experiment has never been reported by other systems including quantum dot system in Ref.~\cite{Kim_2014}. The results can be well explained by the theory of the doubly dressed states. Furthermore, the distance between multimodal structures can be modulated by changing the mean photon number inside the cavity, which is another interesting fluorescence phenomenon modulated by a cavity. Due to the limitation of the cavity-atom coupling strength, the driving field from the cavity in our system is equal to a classical field when the mean photon number inside the cavity is more than one. Therefore, the resonance fluorescence is similar to the emission spectrum reported in Ref.~\cite{He_2015} where the atom is driven by two classcial fields.
    
    By studying the dynamics of the transmon-cavity coupled system, we measure the Rabi oscillations under different classical driving strength and point out its relationship with fluorescence spectra. Moreover, our experiment provides a way to ascertain driven qubits or driven quantum emitters coupling to unwanted degrees of freedom in the environment, i.e., parasitic modes or two-level defects, which is harmful for superconducting quantum computing and quantum network. Besides, we verify that the central component can have subnatural linewidth in this system when $\kappa\ll\gamma$. Our experiment results are consistent with the theoretical models in Ref.~\cite{Freedhoff_1993}. Through our work, the control range of resonant fluorescence by cavity field is expanded, which helps people to have a more in-depth and systematic understanding of resonance fluorescence properties under doubly driven fields.
\begin{acknowledgments}
    Devices were made at the Nanofabrication Facilities at
Institute of Physics in Beijing. Measurements
were performed at Hunan Normal University in Changsha. This work was supported by the National Natural Science Foundation of China under Grant No. 12074117, No. 92365209, Innovation Program for Quantum Science and Technology No. 2021ZD0301800. L.M.K. is supported by the NSFC under Grant Nos. 12247105, 12175060, and 11935006, the  XJ-Lab key project  and the STI Program of Hunan Province under Grant Nos. 23XJ02001 and 2020RC4047.
\end{acknowledgments}

\appendix
	\section{\label{parameters}Measurements of physical parameters of device}
	\subsection{\label{AAsub:level2}Transmon qubit}	
\begin{figure}[h]
    	\includegraphics[scale=0.35]{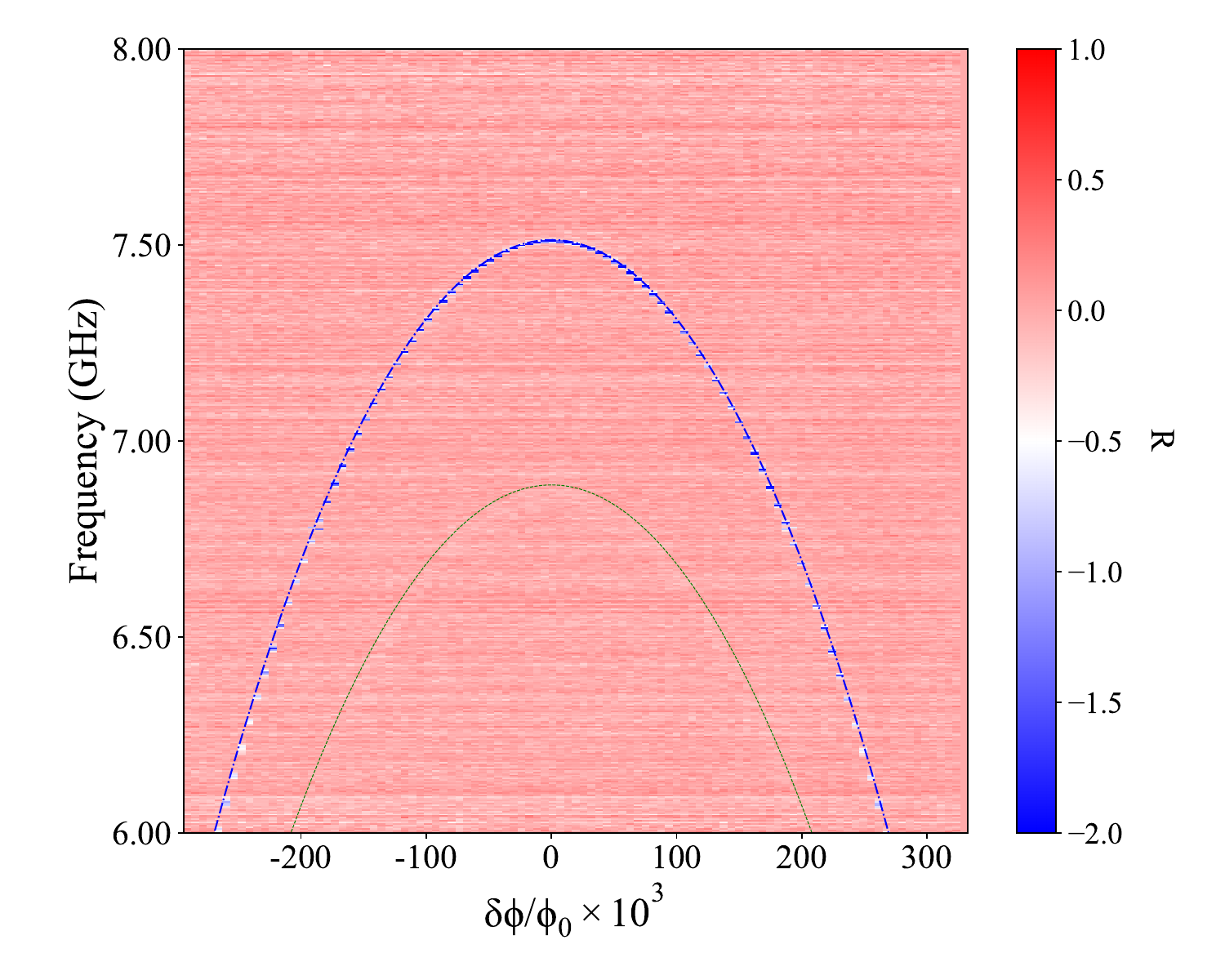}
    	\caption{
     Transmon qubit reflection spectra as a function of biased flux $\delta \phi$ dependence and fitting curves for $E_{01}$ (Blue) and $E_{12}$ (Green) using  Eq. (S1), with $\phi_0=h/2e$.}
     \label{FigS1}
    \end{figure}

	As shown in Fig. 1(a), the transmon qubit consists of a dc-SQUID and a parallel capacitor. The transition frequency from the state $\left| n \right\rangle $ to $\left| n+1 \right\rangle $ of qubit can be expressed as~\cite{Koch_2007}
	\begin{align}\label{S1}
	{{E}_{n,n+1}}=\sqrt{8{{E}_{J}}{{E}_{C}}\cos (\pi \delta \phi /{{\phi }_{0}})}-{{E}_{C}}\left( n+1 \right),
    \end{align} where $\delta \phi$ is the externally biased flux through the SQUID loop, $\phi_0$ is the magnetic flux quanta, $E_J$ and $E_C$ are Josephson energy and charging energy, respectively. In our experiment,  ${{E}_{J}}=13.25$ GHz and ${{E}_{C}}=0.625$ GHz, which are obtained by fitting the energy spectrum in Fig.~\ref{FigS1} with Eq.~(\ref{S1}).  The measured anharmonicity $\alpha \equiv {{E}_{01}}-{{E}_{12}}$ at sweet point is equal to ${{E}_{c}}$.
   
\begin{figure}[h]
	\includegraphics[scale=0.43]{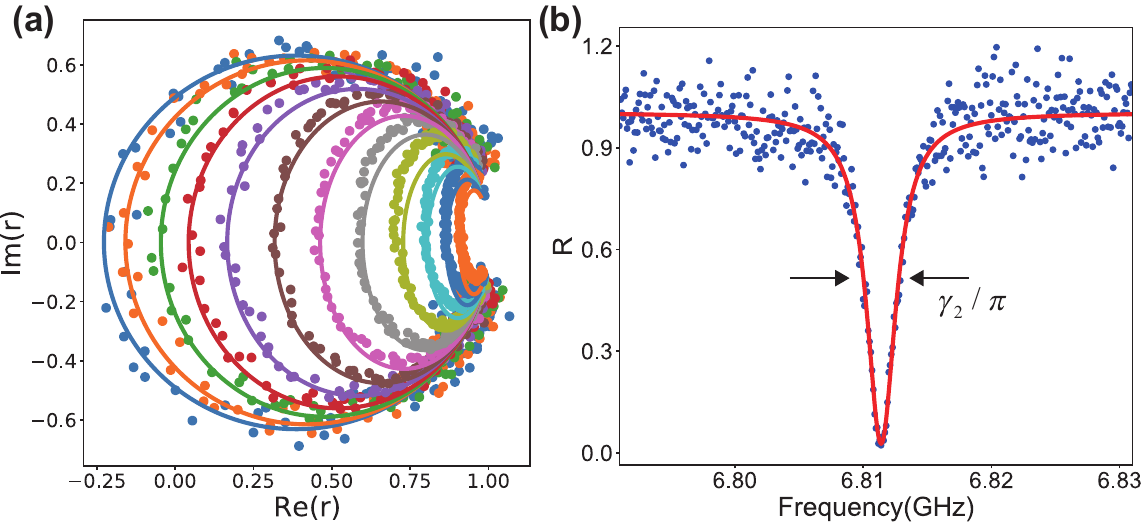}
	\caption{(a) Normalized reflection coefficient ($r_e$) of emission lines for varying probing power with qubit transition frequency at $\omega_a/2\pi$$=$$6.814$ GHz: experimental data (dots) and fitting results (solid lines). (b) Reflection line of the qubit fitted by a Lorentzian line shape. The probing power is $-147\,$dBm.}
 \label{FigS2}
    \end{figure}

    From Refs.~\cite{Astafiev_2010,Peng_2016}, we know that the reflection coefficient of a waveguide coupled with a qubit can be expressed as
    \begin{align}\label{S2}
    {{r}_{e}}=1-\frac{{{\gamma }_{e}}}{\gamma_2 }\frac{1-i\delta \omega /\gamma_2 }{1+{{\left( \delta \omega /\gamma_2  \right)}^{2}}+{{\Omega }^{2}}/\left( {{\gamma }_{1}}{{\gamma_2 }} \right)},
    \end{align} where $\gamma_e$ is the energy emission rate of the qubit to the waveguide. $\gamma_1$ is the incoherent energy relaxation rate, $\gamma_2$ is the total dephasing rate, containing the energy decay rate and the pure dephasing rate. When the probing power is week$\left( \Omega \ll {{\gamma }_{1}},{{\gamma }_{2}} \right)$, ${{r}_{e}}$ can be simplified as
    \begin{align}\label{S3}
    {{r}_{e}}\approx 1-\frac{{{\gamma }_{e}}}{\gamma_2 }\frac{1}{1+i\delta \omega /\gamma_2 },
    \end{align}
that is a circle of radius of ${{{\gamma }_{e}}}/{2\gamma }\;$ on a complex plane. Note that the qubit frequency in our experiment is detuned from cavity, so Eqs.~(\ref{S2}) and (\ref{S3}) can still be used in our system. We measure the reflection lines with probing power from $-149\,$dBm to $-127\,$dBm with a step of $2\,$dB, as shown in Fig.~\ref{FigS2}(a). Fitting the data with Eq.~(\ref{S2}), we find that the emission efficiency of the qubit is $\eta \equiv $${{{\gamma }_{e}}}/{2\gamma_2 }=\;$0.62, when $\Omega \to 0$, which is also the radius of  the largest circle in Fig.~\ref{FigS2}(a).  The peak width in Fig.~\ref{FigS2}(b) corresponds to the total dephasing rate of the atom as $\gamma_2 /2\pi =2.8\,$MHz.
    
    \subsection{\label{Asec:level3}Coupling between the transmon and cavity}
\begin{figure}[h]
    	\includegraphics[scale=0.4]{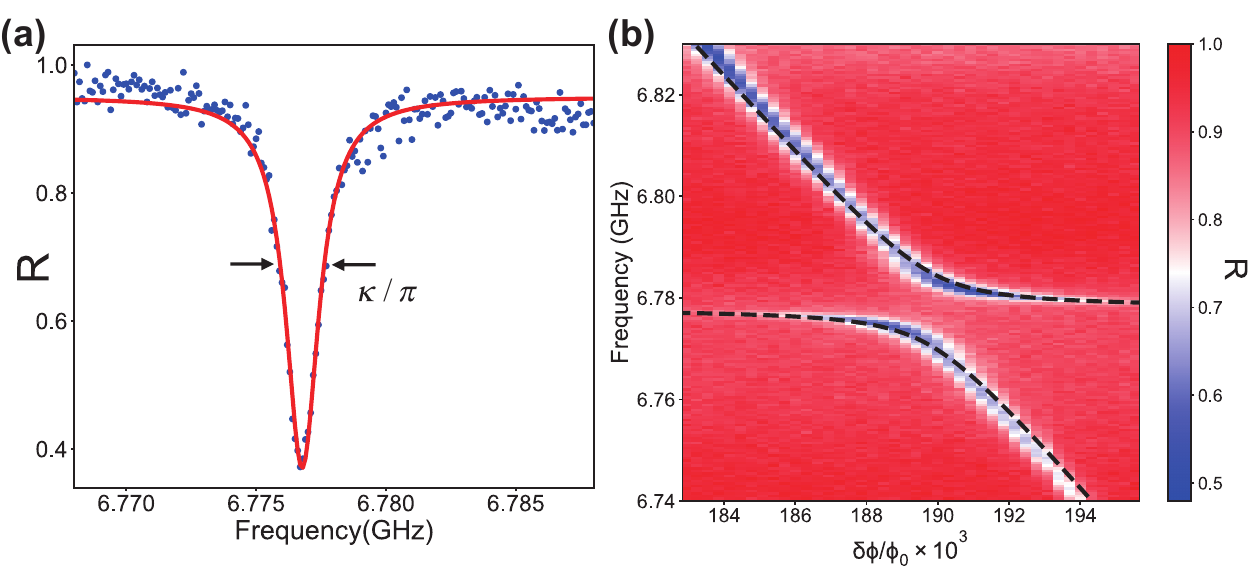}
    	\caption{(a) Reflection line of the cavity fitted by Lorentzian with probing power $-136$ dBm. (b) Reflection spectroscopy of the transmon qubit as a function of $\delta\phi$ near the cavity mode. }
     \label{FigS3}
    \end{figure}

    Figure~\ref{figure1}(a) shows that the other side of the qubit is connected to a coplanar waveguide (CPW) resonator, which exhibits a frequency of 6.777 GHz and a linewidth of $1.5\,$MHz according to the measurements presented in Fig.~\ref{FigS3}(a).
    
    To obtain the interaction strength between the qubit and the resonator, we tune the transition frequency of the qubit to observe the anticrossing in the reflection spectrum as shown in Fig.~\ref{FigS3}(b) and fit it with the Jaynes-Cummings model. In this experiment, we find that the coupling strength is $g_c /2\pi=7.5\pm 1\,$MHz. 

	\section{\label{theory}Caculation of the transmon-cavity coupled system}
        \subsection{\label{AAsec:level0} Effective driving strength for the transmon qubit}
    In our experiments, the driving field is introduced from the cavity side and the driving frequency $\omega_d$ is largely detuned with the cavity. The full Hamitanion of the process under the rotation picture of $\omega_d$ is 
    \begin{align}    
        H=\Delta_c a^{+} a+\frac{\Delta_a}{2} \sigma_z+g_c\left(a^{\dag} \sigma_{-}+a \sigma_{+}\right)+\frac{\Omega_c}{2}\left(a+a^{+}\right),
    \end{align}
    where $\Omega_c$ is strength of the driving field.
    Using the displacement operator as the unitary transformation
    \begin{align}
        U=D(\alpha)=\exp \left(\alpha a^{+}-\alpha^* a\right),
    \end{align}
    and omitting the constant terms, we can obtain 
     \begin{align} \label{S7}
        UHU^{\dag}=&\Delta_c a^{+} a+\frac{\Delta_a}{2} \sigma_z+g_c\left(a^{\dag} \sigma_{-}+a \sigma_{+}\right) \nonumber\\ 
        &-g_c \left(\alpha^*\sigma_{-}+\alpha\sigma_{+}\right)-\Delta_c\left(\alpha a^{+}+\alpha^* a\right)+\frac{\Omega_c}{2}\left(a+a^{+}\right).
    \end{align} 
    When $\alpha=\Omega_c/2\Delta_c$, Eq.~(\ref{S7}) will be simplified as
     \begin{align}     
        UHU^{\dag}=\Delta_c a^{+} a+\frac{\Delta_a}{2} \sigma_z+g_c\left(a^{\dag} \sigma_{-}+a \sigma_{+}\right)+\frac{\Omega}{2}\left(\sigma_{-}+\sigma_{+}\right),
    \end{align}   
    with
    \begin{align}\label{S9}
    \Omega=-2g_c\alpha=-g_c\Omega_c/\Delta_c. 
    \end{align} 
    Therefore, if the input signal is detuned from the cavity resonant frequency, its influence can be equivalent to the atom using Eq.~(\ref{S9}).
        
	\subsection{\label{AAsec:level1} State vectors of the doubly-dressed states}
	We consider the energy splitting ($\hbar\omega_a$) between the ground $\left|g\right\rangle$ and excited $\left|e\right\rangle$ states of the transmon qubit. With a strong driving field  ($\omega_d$, $\Omega$), the effective Hamiltonian of the system is
\begin{align}
	H_{da}=\frac{1}{2}\Delta _{a}\sigma _{z}
	+\frac{\Omega}{2} \left(\sigma _{+}+\sigma _{-}\right),
\end{align}
with $\Delta_{a}=\omega_{a}-\omega_d$. The eigenstates and eigenvalues of $H_{da}$ are
\begin{align}
	H_{da}|\tilde{i},N\rangle =&\left( N\omega _{d}-(-1)^{i}\tilde{\Omega} \right) |\tilde{i},N\rangle,\; i=1,2,\\
	|\tilde{2},N\rangle =&\sin \phi |g,N\rangle -\cos \phi |e,N-1\rangle , \\
|\tilde{1},N\rangle =&\cos \phi |g,N\rangle +\sin \phi |e,N-1\rangle , 
\end{align}
where $\cos ^{2}\phi =\frac{1}{2}-\frac{\Delta a}{2\Omega },\delta =\frac{\Delta_a%
	}{\Omega}$, $\tilde{\Omega} =\frac{\Omega}{2} \left( 1+\delta ^{2}\right)
	^{1/2}$ and $N$ is the mean photon number of the driving field. In our experiments, we let $\Delta_a=0$, so $\cos\phi=1/\sqrt{2}$.
	Considering the dissipative coupling with the cavity ($H_c=\omega_c a^\dag a$), the non-interacting states have degeneracy, when $\omega_c=\omega_d-2\tilde{\Omega}$, as 
\begin{align}	
	E_{|\tilde{1},N-1,n+1\rangle}=E_{|\tilde{2},N,n\rangle}=(N+n)\omega _{d}-(2n+1)\tilde{\Omega} ,
\end{align}
where $n$ is the mean photon number of the cavity and $|\tilde i,N,n\rangle \equiv |\tilde i,N\rangle \otimes |n\rangle$. 

When including the coupling term $\left(W=g_c\left( {{a^\dag }{\sigma _ - } + a{\sigma _ + }} \right) \right)$ between the transmon and the cavity, we can expand $H_{da}+H_c+W$ under the basis vectors $\left[|\tilde{1}, N-n,n\rangle,\;|\tilde{2}, N-n+1,n-1\rangle \right]^T$ as
\begin{align}	
	 & H=H_{da}+H_c+W=\left(%
	\begin{array}{cc}
		E_0 & -g_1 \sqrt{n} \\ 
		-g_1 \sqrt{n} & E_0%
	\end{array}%
\right),
\end{align}
with $g_1=g_c \cos^2 \phi$ and $E_0=N\omega_d-(2n-1)\tilde{\Omega}$. The eigenvalues of $H$ are 
\begin{align}	
	E_{N,\pm n}= N\omega _{d}-(2n-1)\tilde{\Omega} \pm g_{1}\sqrt{n} ,
\end{align}	
with eigenstates which are also called as doubly-dressed states
\begin{align}	
	|N,n\rangle =\frac{1}{\sqrt{2}}\left[|\tilde{2},N-n,n\rangle -|\tilde{1}%
	,N-n+1,n-1\rangle \right], \\
	|N,-n\rangle =\frac{1}{\sqrt{2}}\left[|\tilde{2},N-n,n\rangle +|%
	\tilde{1},N-n+1,n-1\rangle \right].
\end{align}	

\subsection{\label{AAsec:level2} Transition probabilities between doubly-dresssed states}
The transition between doubly-dressed states are realized through operator $\sigma_{+}$ of the transmon and $a^\dag$ of the cavity.
We can expand $\sigma_{+}$ in the picture of doubly-dressed states as
\begin{align}
	\sigma _{+} =& \sum_{n,n^{\prime },N,N^{\prime }}\left\vert
	N,n\right\rangle \left\langle N,n\right\vert \sigma _{+}\left\vert
	N^{\prime },n^{\prime }\right\rangle \left\langle N^{\prime },n^{\prime
	}\right\vert  \nonumber\\
	=&\sum_{n,n^{\prime },N} \left\vert
	N,n\right\rangle  \left\langle N-1,n^{\prime }\right\vert \left\langle N,n\right\vert \sigma _{+}\left\vert
	N-1,n^{\prime }\right\rangle  \nonumber\\	
	=&\sum_{n,n^{\prime },N}\rho _{n,n^{\prime},N}\sigma_{n,n^{\prime },N},\,
(\sigma_{n,n^{\prime },N,N^{\prime }}=0,if\,\,N^{\prime }\neq N-1) 
\end{align}
with $\rho _{n,n^{\prime },N}\equiv\left\vert N,n\right\rangle \left\langle N-1,n^{\prime }\right\vert $ and $\sigma_{n,n^{\prime },N}\equiv \left\langle N,n\right\vert \sigma _{+}\left\vert
N-1,n^{\prime }\right\rangle$.
The transition probabilities caused by $\sigma_{+}$ are
\begin{align} 
\gamma _{\pm n,n^{\prime }} =&\gamma \left\vert \left\langle N,\pm
n\left\vert \sigma _{+}\right\vert N-1,\pm n^{\prime }\right\rangle
\right\vert ^{2}=\gamma \left\vert \sigma_{n,n^{\prime },N}\right\vert ^{2},
\end{align}
and can be detailed as
\begin{equation} \label {eq_gamma}
	\gamma _{\pm n,n^{\prime }}=\left\{ 
	\begin{array}{c}
		\gamma \left\vert \sin \phi \cos \phi \right\vert ^{2},\,\mp n^{\prime
		}= \pm n=0,1,...\\ 
		0,\,\pm n^{\prime }= \pm n=1,2,...\\ 
		\frac{1}{4}\gamma \left\vert \sin \phi \right\vert ^{4} (1+\delta_{n^{\prime},0}),\,n^{\prime }=n-1=0,1,... \\ 
		\frac{1}{4}\gamma \left\vert \cos \phi \right\vert ^{4} (1+\delta_{n,0}),\,n^{\prime }=n+1=1,2,...%
	\end{array}%
	\right..
\end{equation}
We can get the transition probabilities caused by $a^{\dag}$ as
\begin{align}
	\kappa _{\pm n,n^{\prime }} =& \kappa \left\vert \left\langle N,\pm
	n\left\vert a^{\dagger }\right\vert N-1,\pm n^{\prime }\right\rangle
	\right\vert ^{2} \nonumber\\
	=& \left\{ 
\begin{array}{c}
	\frac{1}{\sqrt{2}}, n^{\prime}= \pm (n-1)=0\\
	\frac{1}{2}(\sqrt{n+1}+\sqrt{n}),\,\pm n^{\prime }= \pm (n-1)\neq 0\\ 
	\frac{1}{2}(\sqrt{n+1}-\sqrt{n}),\,\mp n^{\prime }= \pm (n-1)\neq 0
\end{array}%
\right..
\end{align}
Therefore, the total decay rate of state $\vert N,n\rangle$ to lower states is 
\begin{align}
\Gamma _{\pm n}=&\gamma _{\pm n}+\kappa _{\pm n} \nonumber\\
=&\left\{ \begin{array}{c}
	\gamma /2,\,n\neq 0 \\ 
	\gamma \cos ^{2}\phi ,\,n=0%
\end{array}%
\right.+\left\{ \begin{array}{c}
	\frac{\kappa }{2}(2n-1),\,n\neq 0 \nonumber\\ 
	0,\,n=0%
\end{array}%
\right.\\
=&
\left\{ 
\begin{array}{c}
	\frac{\gamma }{2}+\frac{\kappa }{2}(2n-1),\,n\neq 0 \\ 
	\gamma \cos ^{2}\phi ,\,n=0%
\end{array}%
\right.. 
\end{align}

As shown in Ref.~\cite{Freedhoff_1993}, we can find that the populations of state $\vert N,n\rangle$ are
\begin{align}\label{ddpopulation}
	\Pi_{n}=\Pi_0 \prod_{m=1}^n \frac{\gamma \sin ^4 \phi}{\gamma \cos ^4 \phi+(2 m-1) \kappa},\, n=1,2,...
\end{align}
and $\Pi_{-n}=\Pi_{n}$, where $\Pi_0$ is the normalization coefficient and represents the populations of state $\vert N,0\rangle$.

\subsection{\label{AAsec:level3} Asymmetry of the flourescence spectrum}
Using the quantum regression theorem,  the steady-state emission spectrum of the transmon can be simplified as  
\begin{widetext}
\begin{align}
	S(\omega ) =  \pi ^{-1} \lim_{t\rightarrow \infty }\left[\text{Re} \int_{0}^{%
		\infty }e^{-i\omega \tau }\left\langle \sigma _{+}(t+\tau )\sigma
	_{-}(t)\right\rangle d\tau \right]  
	= \pi ^{-1} \text{Re} \int_{0}^{\infty }e^{-i\omega \tau } \sum_{n,n^{\prime },N}\sigma _{n,n^{\prime },N}\left\langle \rho
	_{n,n^{\prime },N}(\tau )\sigma _{-}\right\rangle _{s} d\tau .
\end{align}
According to the derivation process of Ref.~\cite{Freedhoff_1993}, the spectra around $(\omega _{a}\pm\Omega_0)$ are

\begin{align}
	S^{(1)}\left( \omega -\omega _{a}-\Omega_0 \right) =\frac{1}{\pi\gamma }%
	\sum_{n=0,\pm }^{\infty }\frac{\gamma _{n,n+1}\Pi _{n}\Gamma _{\pm (%
			\sqrt{n+1}\pm \sqrt{n})}}{\left[ \omega-\omega _{a}-\Omega_0 \pm g_{1}(\sqrt{%
			n+1}\pm \sqrt{n})\right] ^{2}+\Gamma _{\pm (\sqrt{n+1}\pm \sqrt{n})}},\\
	S^{(-1)}\left( \omega -\omega _{a}+\Omega_0 \right) = \frac{1}{\pi\gamma}%
	\sum_{n=0,\pm }^{\infty }\frac{\gamma _{n+1,n}\Pi _{n+1}\Gamma _{\pm (%
			\sqrt{n+1}\pm \sqrt{n})}}{\left[ \omega-\omega _{a}-\Omega_0 \pm g_{1}(\sqrt{%
			n+1}\pm \sqrt{n})\right] ^{2}+\Gamma _{\pm (\sqrt{n+1}\pm \sqrt{n})}},
\end{align}
with the linewidth $\Gamma _{\pm (\sqrt{n+1}\pm \sqrt{n})}=\gamma/2+n\kappa$ for $n=1,2,...$ or $\frac{\gamma}{2}(\frac{1}{2}+\sin^2\phi)+\frac{\kappa}{4}$ for n=0.

For the experimental results presented in Fig. 2, the intensity of $S^{(1)}\left( \omega -\omega _{a}-\Omega_0 \right)$ depends on $\gamma _{0,1}\Pi _{0}\Gamma _{\pm 1}$, while $S^{(1)}\left( \omega -\omega _{a}+\Omega_0 \right)$ depends on $\gamma _{1,0}\Pi _{1}\Gamma _{\pm 1}$. Because $\Delta=0$ in our experiment, $\gamma _{1,0}=\gamma _{0,1}=\gamma/8$ according to Eq.~(\ref{eq_gamma}). However, the spectra intensity around $\omega _{a}-\Omega_0$ is smaller because $\Pi _{1}=\Pi _{0}\gamma/(\gamma+2\kappa)\textless \Pi _{0}$.
\end{widetext}

\section{\label{Asec3:level3}Measurement Setup}
   \begin{figure}[h] 
    \centerline{\includegraphics[scale=0.45]{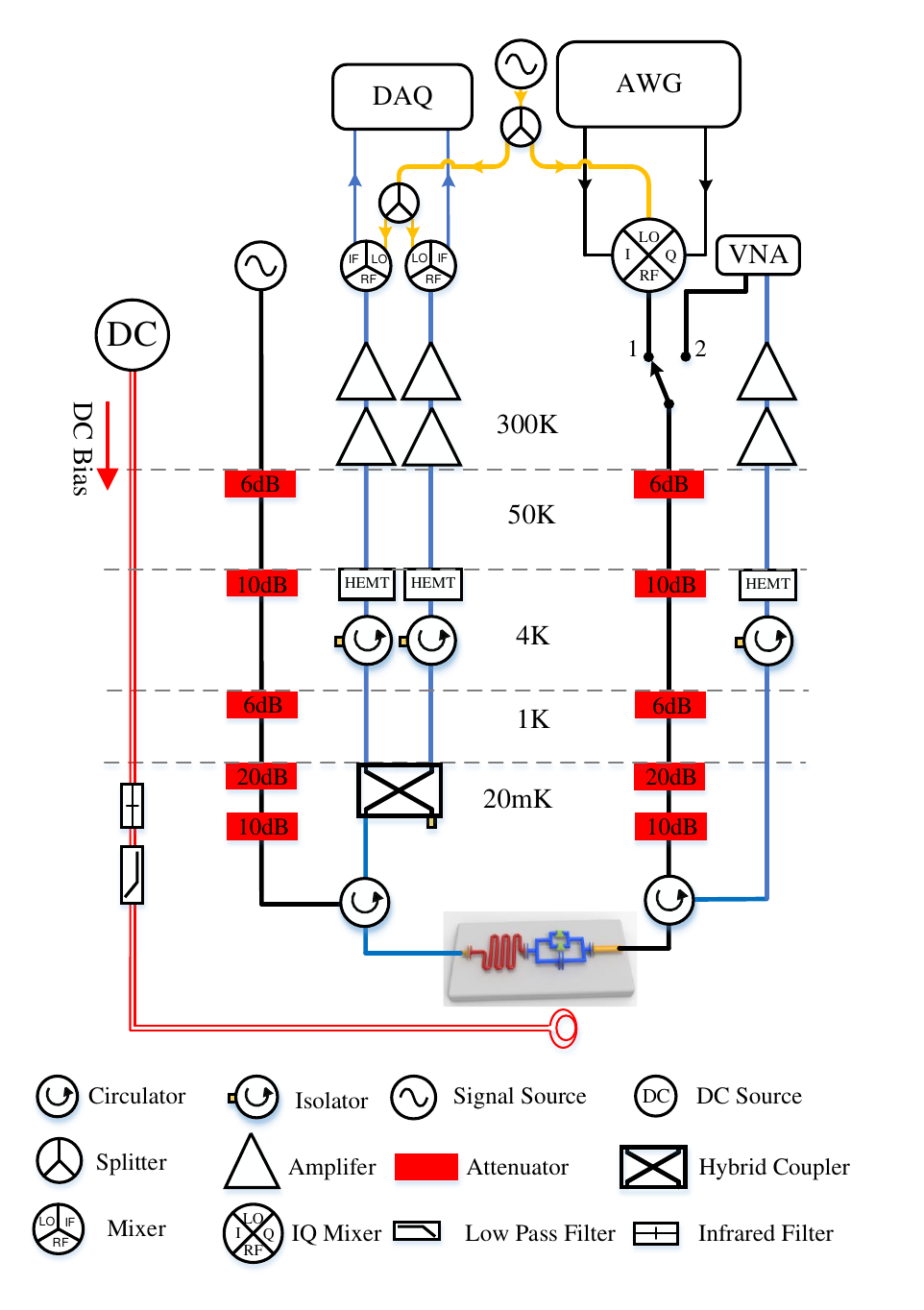}}
	\caption{Schematic diagram of the measurement setup.}
	 \label{circuits}
\end{figure}  
As shown in Fig.~\ref{circuits}, our time-domain measurement uses the heterodyne demodulation setup. The input signal is generated with single-sideband (SSB) modulation technique, which up-coverts the 80$\,$MHz signals from an arbitrary wave generator (AWG) to the transition frequency of the qubit. The output signal from the device is transmitted through a circulator and mixed with vacuum state by using a $90^{\circ}$ hybrid coupler at 20$\,$mK stage. Then, the mixed signal is split into two signals and transmitted in two channels with nominal identical gain, respectively. The signals in the two channels are amplified by cryo-amplifiers (HEMT) at $4\,$K stage and amplifiers at room temperature. Then, we can detect the correlation function of the output signal emitted from the device or evolution dynamics of the device using this Hanbury Brown-Twiss (HBT) type measurement. Before the amplified signals are collected by the data acquisition (DAQ) card with 400$\,$MHz  sampling rate, they are down-converted to 80$\,$MHz. The subsequent calculation of the correlation function and other parameters are carried out by GPU in real time. Measurement of device parameters in Appendix~\ref{parameters} is done by a vector network analyzer (VNA), when the single-pole dual switcher in Fig.~\ref{circuits} is at position~2.   
	
\section{\label{Asec5:level5}Linewidth of central peak}
  \begin{figure}[h]
	\centerline{\includegraphics[scale=0.45]{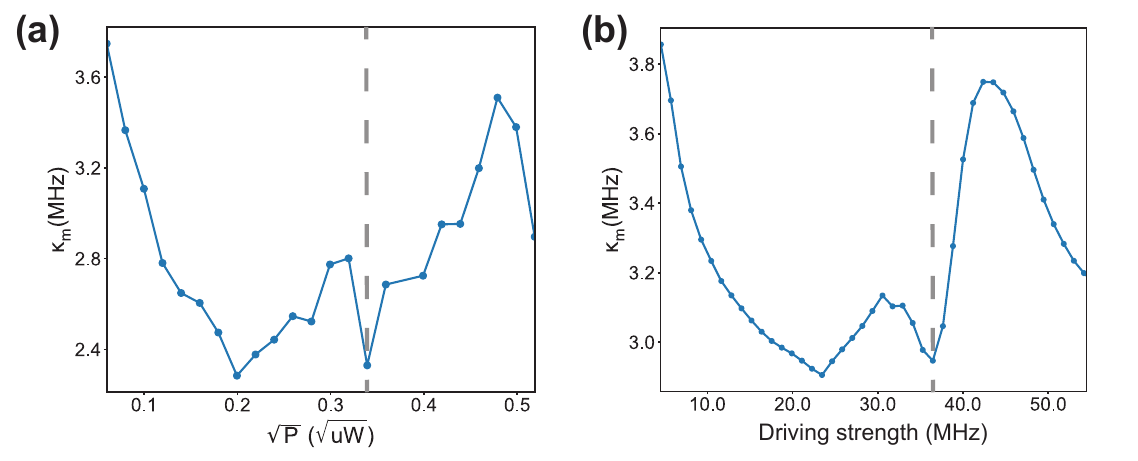}}
	\caption{(a) Linewidth of central peaks against $\sqrt{P}$ in the experiment. The fittings of peaks are done by Lorentzian curves. (b) Linewidth of central peaks from simulation data. The gray dash line is the driving amplitude that cause the doubly dressed state. }
	\label{S6}
\end{figure}
In Fig. 4(e-f), we can see that the energy relaxation time is extended at doubly dressed situation, which can also be shown by analysing the linewidth of the central peak in spectrum. The linewidth has a minimum value, when the system is in doubly dressed state as shown in Fig.~\ref{S6}. The trends of theoretical and experimental results are similar. The changes of linewidth are caused by the dressing effect from the cavity. This effect shows up, when the driving strength is close to the detuning between the qubit and the cavity. However, the decay rate of cavity in our experiment is only about 46 percent smaller than that of the qubit, so the decrease of central peak linewidth is not very obvious.

    \bibliographystyle{apsrev4-2}
	\bibliography{refrence_fluo}
	\end{document}